\documentclass[prd,twocolumn,showpacs,preprintnumbers,amsmath,amssymb,superscriptaddress]{revtex4-1}

\usepackage{graphicx}
\usepackage{bm}
\usepackage{amsfonts}
\usepackage{CJK}

\def\fm {\mathop{\hbox{fm}}}
\def\MeV {\mathop{\hbox{MeV}}}

\def\Re {\mathop{\hbox{Re}}}
\def\Sign {\mathop{\hbox{Sign}}}

\def\DU  {\mathop{{\cal D}\hbox{U}}}

\def\dd  {\mbox{d}}
\newcommand\detn[1]{\mbox{det}_{#1}}

\newcommand{\beq}{\begin{equation}}
\newcommand{\eeq}{\end{equation}}
\newcommand{\beqa}{\begin{eqnarray}}
\newcommand{\eeqa}{\end{eqnarray}}

\newcommand\comment[1]{}

\begin{document}

\preprint{UK/11-02}

\begin{CJK*}{UTF8}{} 
\title{Critical point of $N_f = 3$ QCD from lattice simulations in the canonical ensemble}
\CJKfamily{gbsn}
\author{Anyi Li(李安意)}
\email{anyili@phy.duke.edu}
\affiliation{Department of Physics, Duke University, Durham, North Carolina 27708, USA}
\author{Andrei Alexandru}
\email{aalexan@gwu.edu}
\affiliation{Physics Department, The George Washington University, Washington DC 20052, USA}
\author{Keh-Fei Liu(刘克非)}
\email{liu@pa.uky.edu}
\affiliation{Department of Physics and Astronomy, University of Kentucky, Lexington, Kentucky 40506, USA}

\collaboration{$\chi$QCD Collaboration}

\begin{abstract}
A canonical ensemble algorithm is employed to study the phase diagram of $N_f = 3$ QCD
using lattice simulations. We lock in the desired quark number sector using an exact Fourier
transform of the fermion determinant. We scan the phase space below $T_c$ and look for an S-shape
structure in the chemical potential, which signals the coexistence phase of a first order phase
transition in finite volume. Applying Maxwell construction, we determine the boundaries of the
coexistence phase at three temperatures and extrapolate them to locate the critical point.
Using an improved gauge action and improved Wilson fermions on lattices with a
spatial extent of $1.8 \fm$ and quark masses close to that of the strange,
we find the critical point at $T_E = 0.925(5)~T_c$ and baryon
chemical potential $\mu_B^E = 2.60(8)~T_c$.

\end{abstract}


\maketitle
\end{CJK*}

QCD is expected to have a rich phase diagram at finite temperature and finite density. Current lattice
calculations have shown that the transition from the hadronic phase to QGP phase is a rapid
crossover~\cite{Aoki:2009sc,Karsch:2008fe}.
For large baryon chemical potential and very low temperature, a number of models suggest that the transition is
a first order. If this is the case, when the chemical potential is lowered and temperature raised,
this first order phase transition is expected to end as a second order phase transition point --- the critical
point.  However, lattice QCD
simulations with chemical potential are difficult due to the notorious ``sign problem''.
The majority of current simulations
are focusing on small chemical potential region $\mu_q/T \ll 1$ where the ``sign problem'' appears to be
under control.
Up to now, all the $N_f = 3$ or 2+1 simulations are based on the grand canonical ensemble ($T$, $\mu_B$ as
parameters) with staggered fermions.
The results from the multi-parameter reweighting~\cite{Fodor:2004nz}, Taylor expansion with
small $\mu$~\cite{Allton:2005gk, Gavai:2008zr} and the curvature of the critical
surface~\cite{deForcrand:2008vr} are not settled and  need to be cross-checked. Even the existence of the critical point
is in question~\cite{deForcrand:2008vr}. We employ an
algorithm, which is not restricted to small chemical potential because of the mitigation of the sign problem under the current parameter settings, to study this problem.

\begin{figure}[b]
  \center
\includegraphics[width=2.0in]{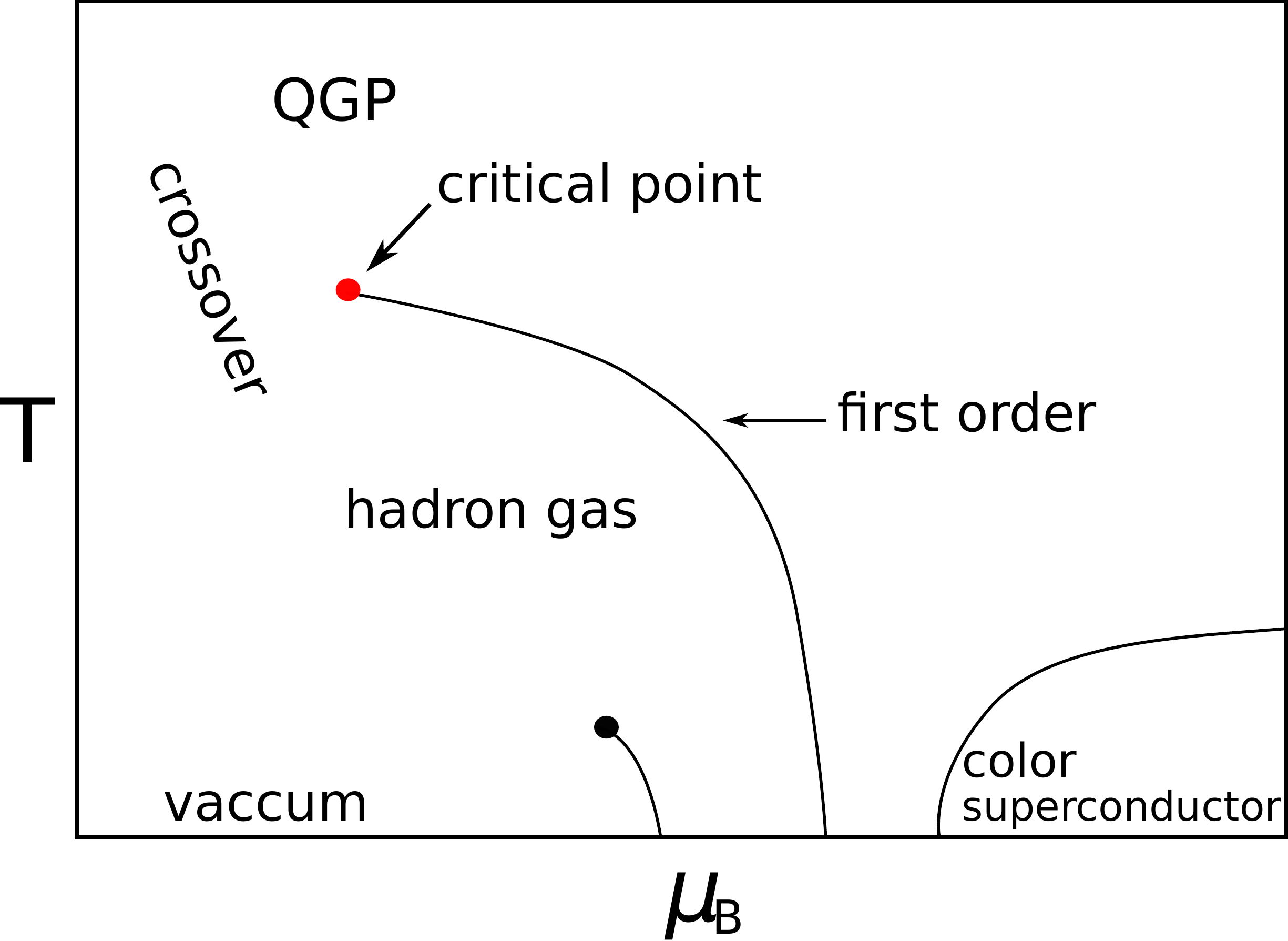}
\caption{Conjectured QCD phase diagram}\label{fig:qcd-phase-diagram}
\end{figure}

In this letter, we adopt an exact Monte Carlo algorithm~\cite{Liu:2003wy, Alexandru:2005ix, Alexandru:2007bb}
based on the canonical partition function~\cite{Engels:1999tz, Liu:2000dj, Liu:2002qr, Azcoiti:2004ri, deForcrand:2006ec, Kratochvila:2005mk}
which is designed to alleviate the determinant fluctuation problems.
As it turns out, the sign fluctuations are not serious on the lattices used in the present study, as we shall see later.
In the canonical ensemble simulations in finite volume, the coexistence phase of a
first order phase transition has a characteristic S-shape as a
function of density due to the surface tension. This finite-volume property has
been exploited successfully to identify the phase boundaries via the Maxwell construction in studies
of phase transition with the staggered fermions~\cite{deForcrand:2006ec, Kratochvila:2005mk} and
clover fermions~\cite{Li:2010qf} for the $N_f=4$ case which is known to have a first order phase transition at $\mu =0$.
In these benchmark studies the
boundaries were identified
at three temperatures below $T_c$, and they were extrapolated in density and
temperature to show that the
intersecting point indeed coincides with the independently identified first order transition point
at $T_c$ and $\mu =0$~\cite{Li:2010qf}. In view of the success of the
$N_f =4$ study, we extend this method to the more realistic
$N_f = 3$ case~\cite{Li:2009ju,Li:2010dya}. Although the real world contains two light quarks and
one heavier strange quark, the three degenerate flavor case has a
similar phase structure. Our primary goal in this study
is to determine whether a first order phase transition exists
for $N_f=3$ and where the critical point is located.

With the aid of recently developed matrix reduction
technique~\cite{Alexandru:2010yb, Nagata:2010xi, Gattringer:2009wi}, we scan the chemical potential
as a function of baryon number for four temperatures below $T_c$ which is determined at zero chemical potential,
and we observe  clear signals for a first order phase transition for temperatures below $0.93~T_c$.
The phase boundaries of the coexistence phase are determined and then extrapolated in temperature and density
to locate the critical point at $T_E = 0.927(5)~T_c$ and $\mu_B^E = 2.60(8)~T_c$. Our results are based on
simulations on $6^3\times 4$ lattices with clover fermion action with quark masses which correspond to the
pion mass from $750 \MeV$ for the lowest temperature to $775 \MeV$ for the highest temperature.

The canonical partition function in lattice QCD can be derived from the fugacity expansion of the
grand canonical partition function,
\beq
Z(V,T,\mu) = \sum_{k} Z_C(V, T, k) e^{\mu k/T}, \label{eq:fugacity}
\eeq
where $k$ is the net number of quarks (number of quarks minus the number of anti-quarks) and $Z_C$
is the canonical partition function. Using the fugacity expansion, it can be shown that the canonical
partition function can be written as a Fourier transform of the grand canonical
partition function,
\beq
\label{eq:Z_C}
Z_C(V, T, k) =
\frac{1}{2\pi} \int_0^{2\pi} \mbox{d}\phi \,e^{-i k \phi} Z(V, T,\mu)|_{\mu=i\phi T},
\eeq
upon introducing an imaginary chemical potential $\mu = i \phi T$. After integrating out the fermionic part
in Eq.~(\ref{eq:Z_C}), we get an expression
\beq
Z_C(V, T, k) = \int \DU e^{-S_g(U)} \detn{k} \label{eq:canonical}
M^{N_f}(U),
\eeq
where
\beq
\detn{k} M^{N_f}(U) \equiv \frac{1}{2\pi}\int_0^{2\pi} \dd\phi\,e^{-i k \phi} \det M(m,
\phi;U)^{N_f} , \label{eq:detk}
\eeq
is the projected determinant for the fixed net quark number $k$. $N_f$ is the number of flavors. We shall
use the recently developed matrix reduction technique to compute the projected determinant
exactly~\cite{Alexandru:2010yb}.

Using charge conjugation symmetry, one can show that $\detn{k} M^{N_f}(U)$ is real, but not necessarily positive.
Due to the sign fluctuation, there can potentially be a sign problem at large quark number and low
temperature. For more detailed discussion about the properties of the canonical ensemble, we refer the
reader to Ref.~\cite{Li:2010qf}. To simulate Eq.~(\ref{eq:canonical}) dynamically,
we rewrite canonical partition function as
\beqa
\! \!\! Z_C(V, T, k) &= &\int \DU e^{-S_g(U)} \mbox{det}M^{N_f}(U)W(U)\alpha(U),
\eeqa
where
\beqa
W(U) &=& \frac{\mathop{|\Re}\detn{k}M^{N_f}(U)|}{\mbox{det}M^{N_f}(U)}, \nonumber \\
\alpha(U) &=& \Sign(\mathop{\Re}\detn{k}M^{N_f}(U)). \label{eq:sign_def}
\eeqa
Our strategy to generate an ensemble is to employ Metropolis accept/reject method based on
the weight $W(U)$ and fold the phase factor $\alpha(U)$ into the measurements. In short, during
the simulation, the candidate configuration is ``proposed'' by the standard Hybrid Monte Carol
algorithm and then an accept/reject step is used for the correct probability. Note the two-step
simulation with HMC and accept/reject based on $W(U)$ reduces the fluctuation problem~\cite{Alexandru:2005ix}
and accept/reject step based on the exact projected determinant $\detn{k} M^{N_f}(U)$ ensures
that the simulation remains in the specific canonical sector with quark number
$k \neq 0$.

The lattice spacing and the pion mass are determined by using dynamically generated ensembles on
$12^3\time 24$ lattices for each $\beta$. To locate the pseudo critical temperature $T_c$, we
varied $\beta$ to look for the peak of the Polyakov loop susceptibility. We run simulations for
five different volumes ($6^3, 8^3, 10^3, 12^3, 16^3 \times4$) and found that the peak of the
susceptibility hardly depends on the volume.  This is consistent with the finding on large volumes and
physical quark masses that the finite temperature transition for the $N_f=3$ case is a crossover
at zero chemical potential~\cite{Aoki:2009sc,Karsch:2008fe}.

To determine the location of the phase transition at non-zero baryon density, we pick four temperatures
below $T_c$ ($0.85~T_c$, $0.87~T_c$, $0.90~T_c$, $0.93~T_c$) and vary the net quark number from 3 to 54
in steps of 3 (for fractional baryon number the partition function
vanishes). This corresponds to the baryon number $n_B$ from 1 to 18
and a density between that of the nuclear matter and 18 times of that. The chemical potential is calculated
and plotted as a function of the net baryon number $n_B$. In the canonical ensemble, the baryon chemical
potential is calculated by taking the difference of the
free energy after adding one baryon, i.e.
\beqa
\left<\mu\right>_{n_B} &= &\frac{F(n_B+1)-F(n_B)}{(n_B+1)-n_B} = -\frac{1}{\beta}\ln \frac{\left<\gamma(U)\right>_o}
{\left< \alpha(U)\right>_o}
\label{eq:baryon chemical potential}
\eeqa
where
\beq
\gamma(U)= \frac{{\Re\,}\detn{3n_B+3} M^{n_f}(U)} {\left|{
\Re\,}\detn{3n_B} M^{n_f}(U)\right|}.
\eeq
is measured in the ensemble with $n_B$ baryon number
and $\left<\right>_o$ in Eq.~\eqref{eq:baryon chemical potential} stands for the average over the ensemble
generated with the measure $\left|\Re \detn{3n_B}M^{n_f}(U)\right|$.

\begin{figure}[t]
\centering
\includegraphics[width=1.65in]{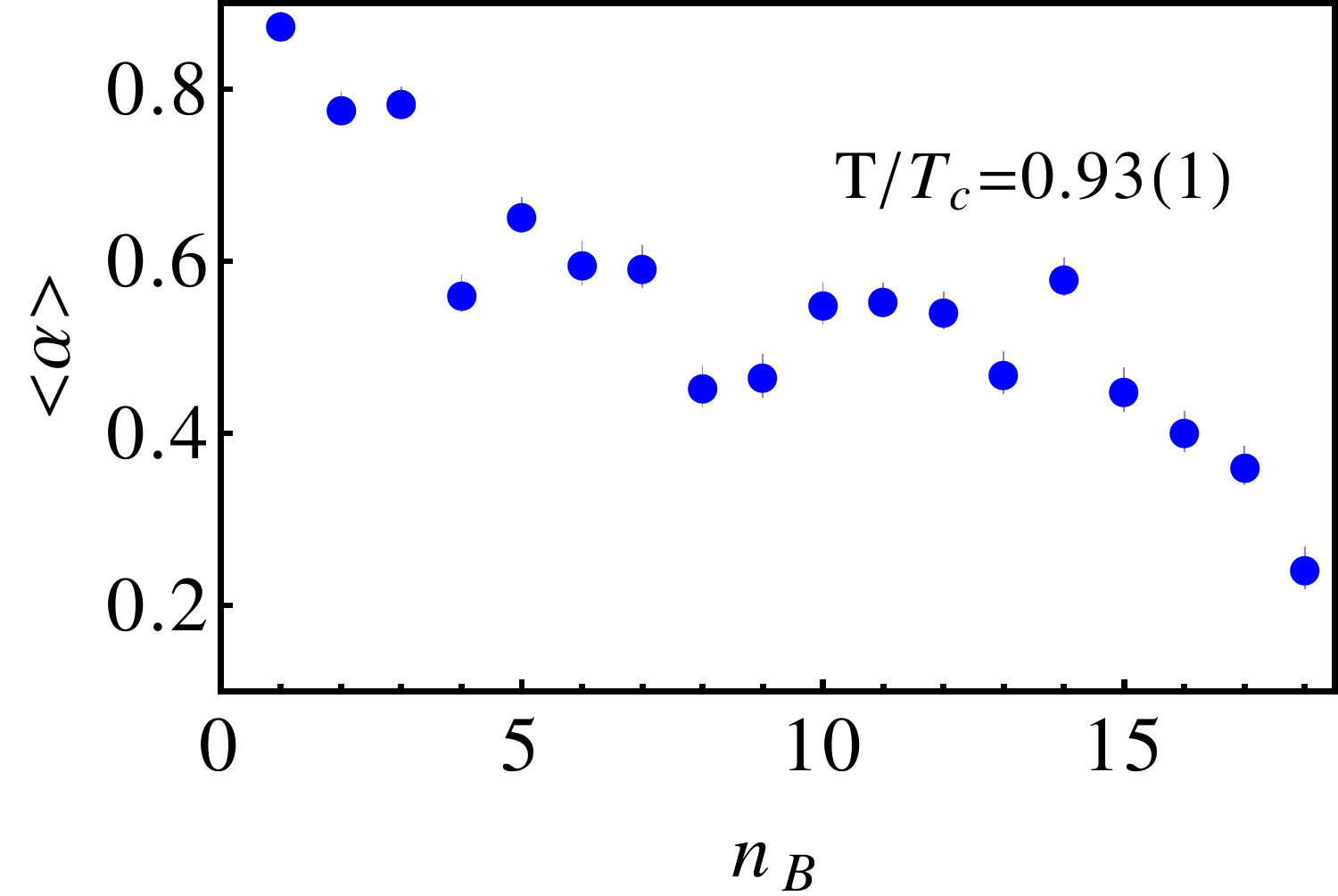}
\includegraphics[width=1.65in]{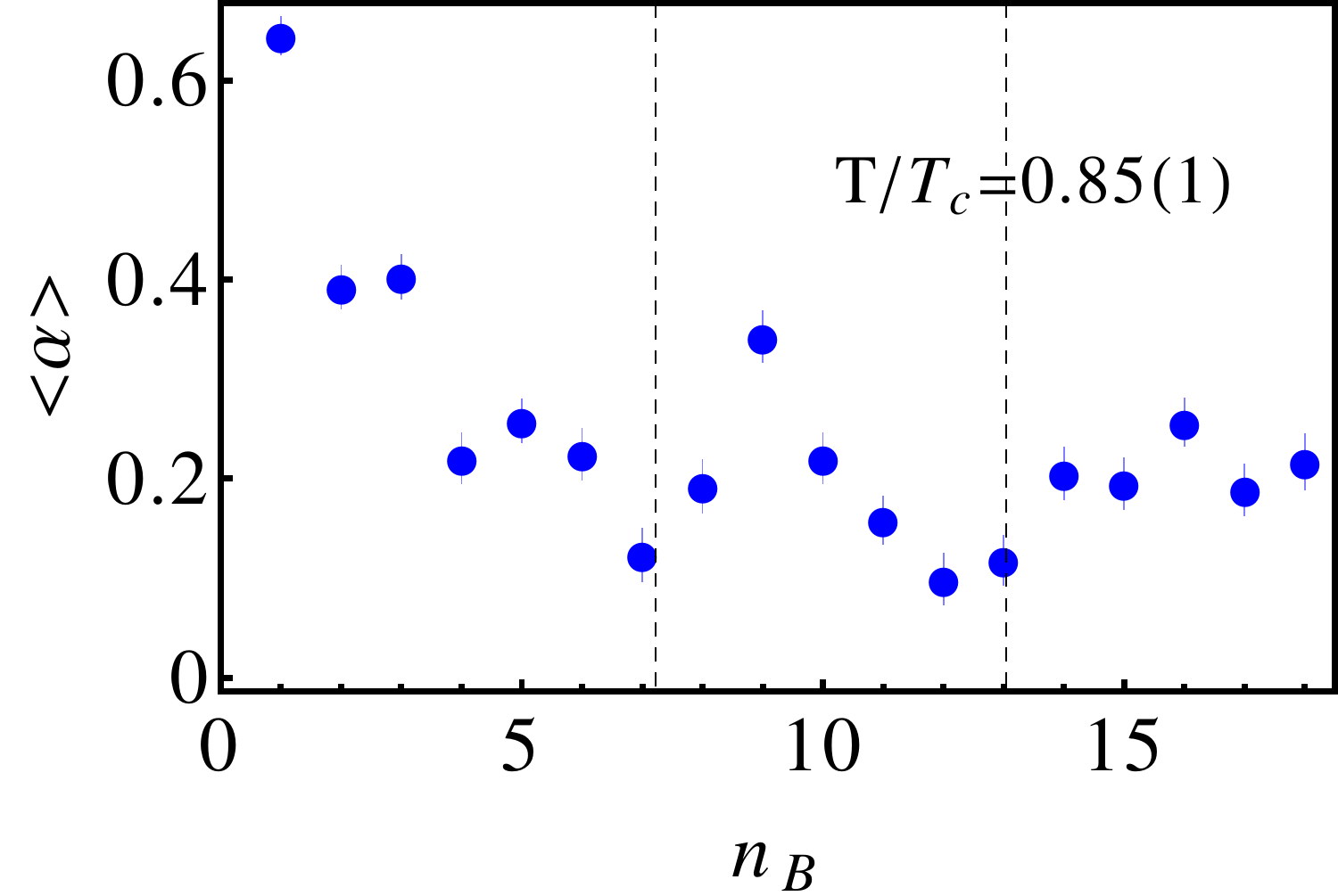}
\caption{Average sign as a function of $n_B$ for highest and lowest temperatures ($0.93~T_c$ and $0.85~T_c$) used in this study. Dashed
lines represent the phase boundaries of the coexistence phase.}\label{fig:sign}
\end{figure}

\begin{figure*}[t]
\includegraphics[width=2.0in]{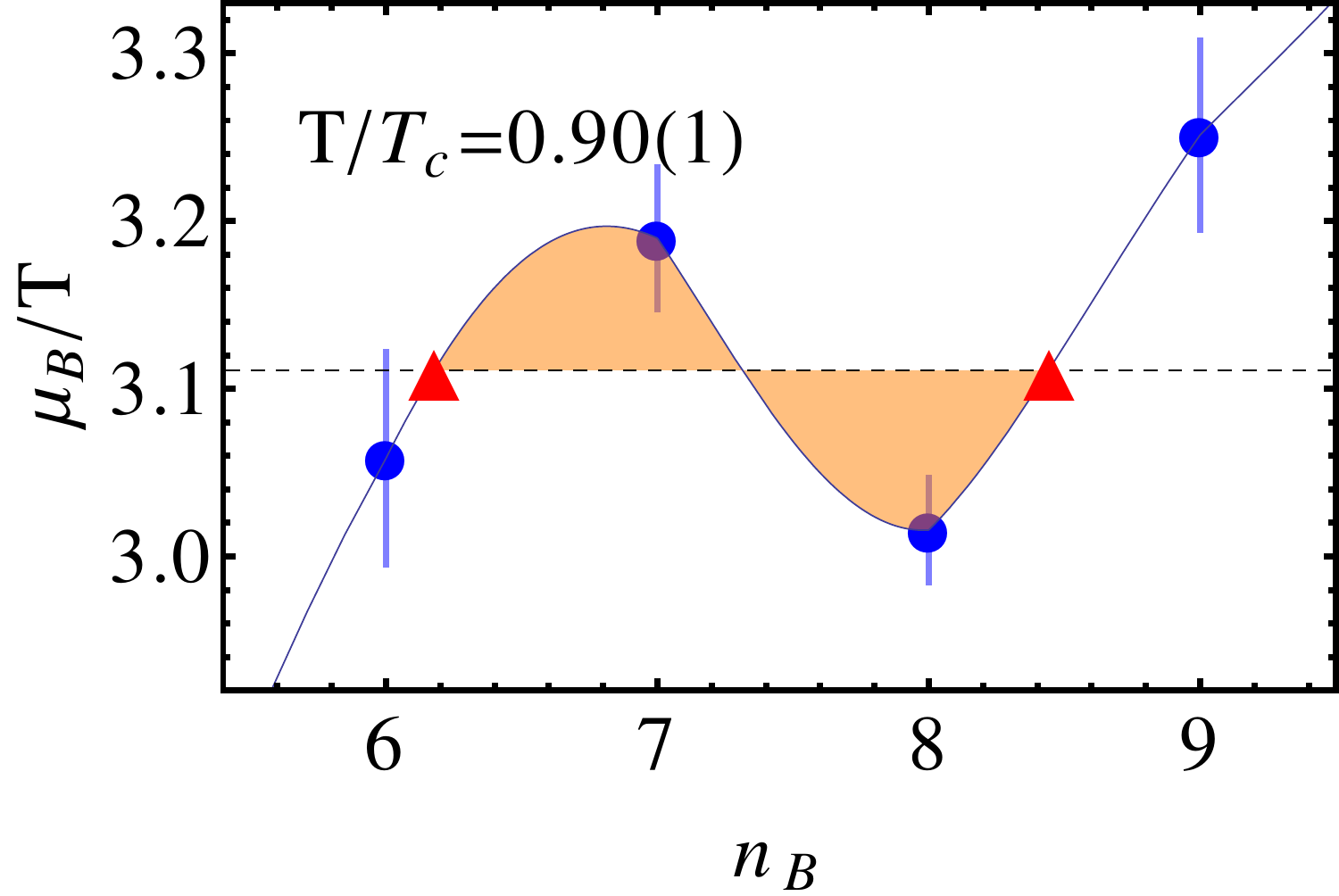}
\includegraphics[width=2.0in]{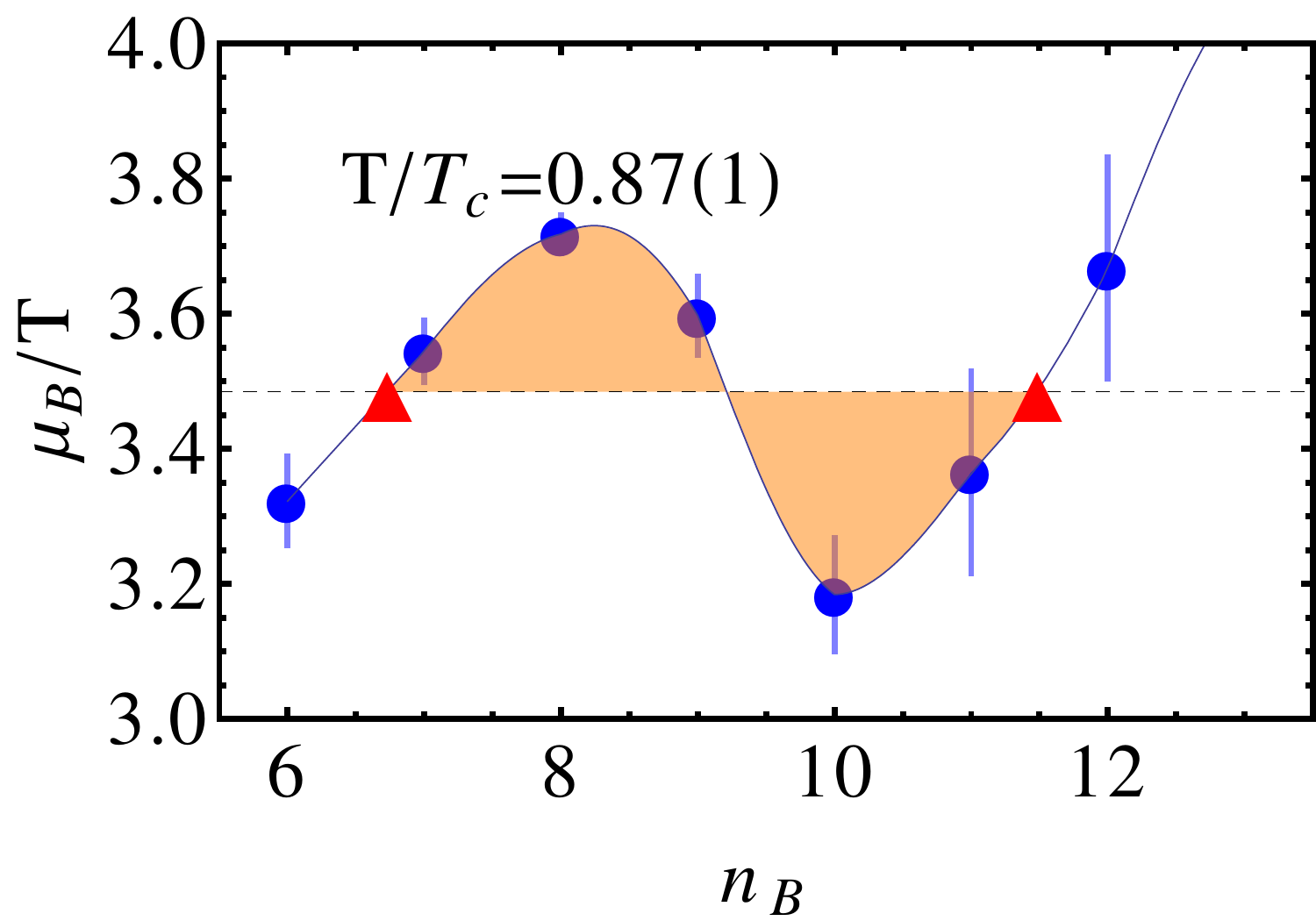}
\includegraphics[width=2.0in]{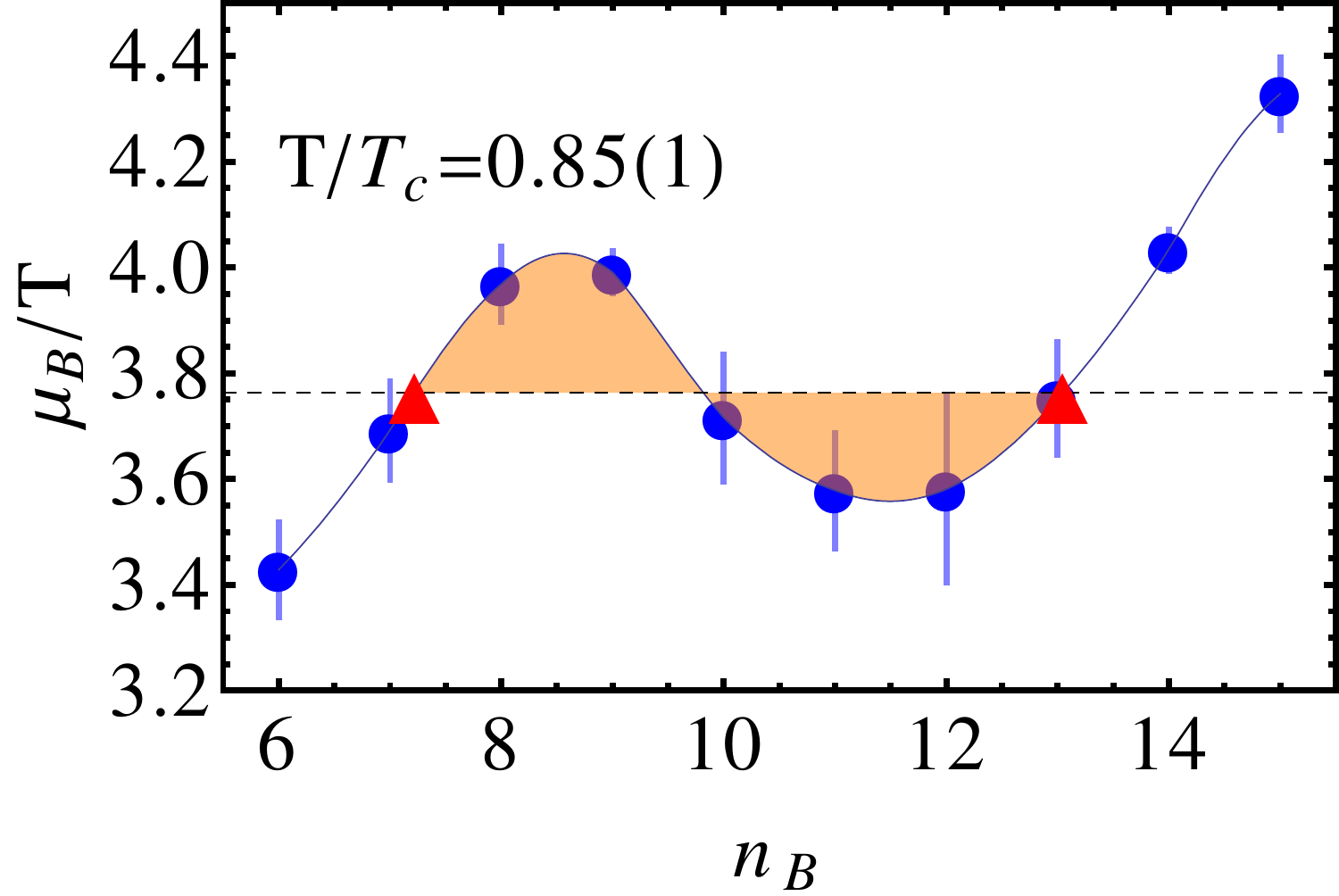}
\caption{Maxwell constructions for $T=0.90~T_c$, $T=0.87~T_c$ and $T=0.85~T_c$ with the horizontal dashed
line indicating the constant $\widetilde{\mu_B}/T$ and red triangles indicating the mixed phase boundaries
at $n_{B_1}$ and $n_{B_2}$.}\label{fig:maxwell}
\end{figure*}

As a first check, we examine the magnitude of the sign fluctuations. The
average sign in Eq.~(\ref{eq:sign_def}) appears in the denominator of Eq.~(\ref{eq:baryon chemical potential})
and can lead to a sign problem when its error bars overlap with zero. This quantity is
plotted in Fig.~\ref{fig:sign} for the highest and lowest temperatures. We see that all of them are more than $3 \sigma$ above zero. This result is better than the previous ones
based on the winding number expansion method~\cite{Li:2010qf,Li:2010dya}, presumably due to the adoption of
the exact projection of the determinant~\cite{Alexandru:2010yb}.
Thus, we believe that the sign fluctuations are not a problem for this study.

\begin{figure}[b]
\includegraphics[width=2.0in]{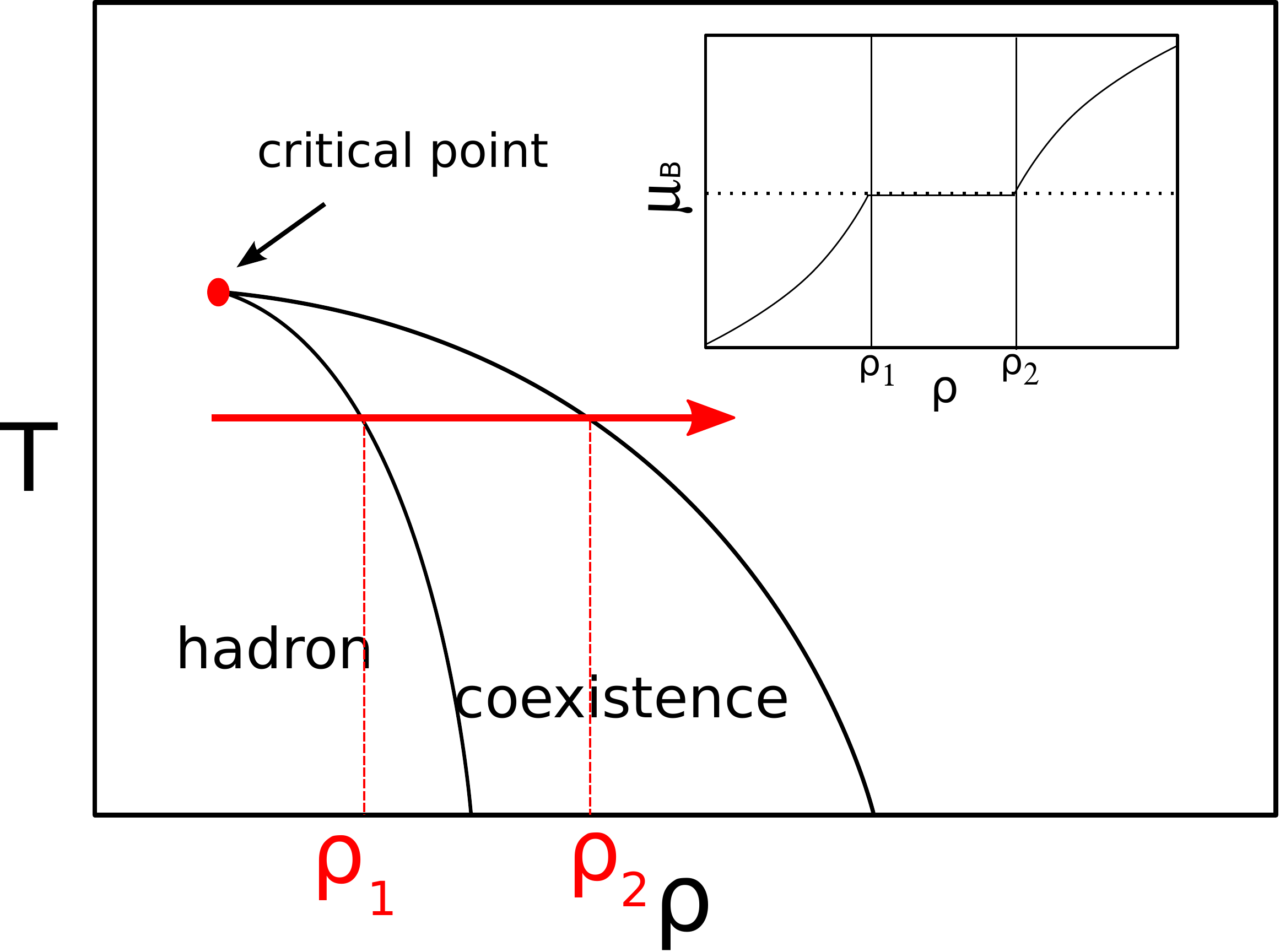}
\caption{Schematic plot illustrating the scanning we use to locate
the boundaries of the mixed phase for QCD with \mbox{$N_f=3$.} The infinite volume expectation for chemical
potential as a function of density is shown in the inset.}\label{fig:t_rho_scan}
\end{figure}

We would like to point out the difference between the phase diagram in the grand canonical ensemble and
the one in the canonical ensemble. We plot the expected canonical ensemble phase diagram in
Fig.~\ref{fig:t_rho_scan} in contrast to that in the grand canonical ensemble in Fig.~\ref{fig:qcd-phase-diagram}.
The first order phase transition line in the grand canonical $T - \mu$ diagram
becomes a phase coexistence region in the $T$ - $\rho$ diagram of the canonical ensemble, which has
two boundaries that separate it from the pure phases. The two boundaries will eventually meet at one
point. This point is the critical point at nonzero baryon chemical potential.

Once one enters the coexistence region in a finite volume, the contribution from the surface tension
causes the appearance of a ``double-well'' in the effective free energy whose derivative with respect
to density leads to an S-shaped behavior in the chemical potential versus baryon number plot~\cite{Ejiri:2008xt}.
However, in the thermodynamic limit, the surface tension contribution goes away since it is a surface
term while the free energy scales with the volume; the chemical potential will then stay constant
in the coexistence phase region. The behavior of the baryon chemical potential in the thermodynamic
limit is shown as an inset in Fig.~\ref{fig:t_rho_scan}. $\rho_1$ and $\rho_2$
mark the lower and upper boundaries of the coexistence phase at a given temperature below $T_c$.

\begin{figure}[b]
\centering
      \includegraphics[width=1.5in]{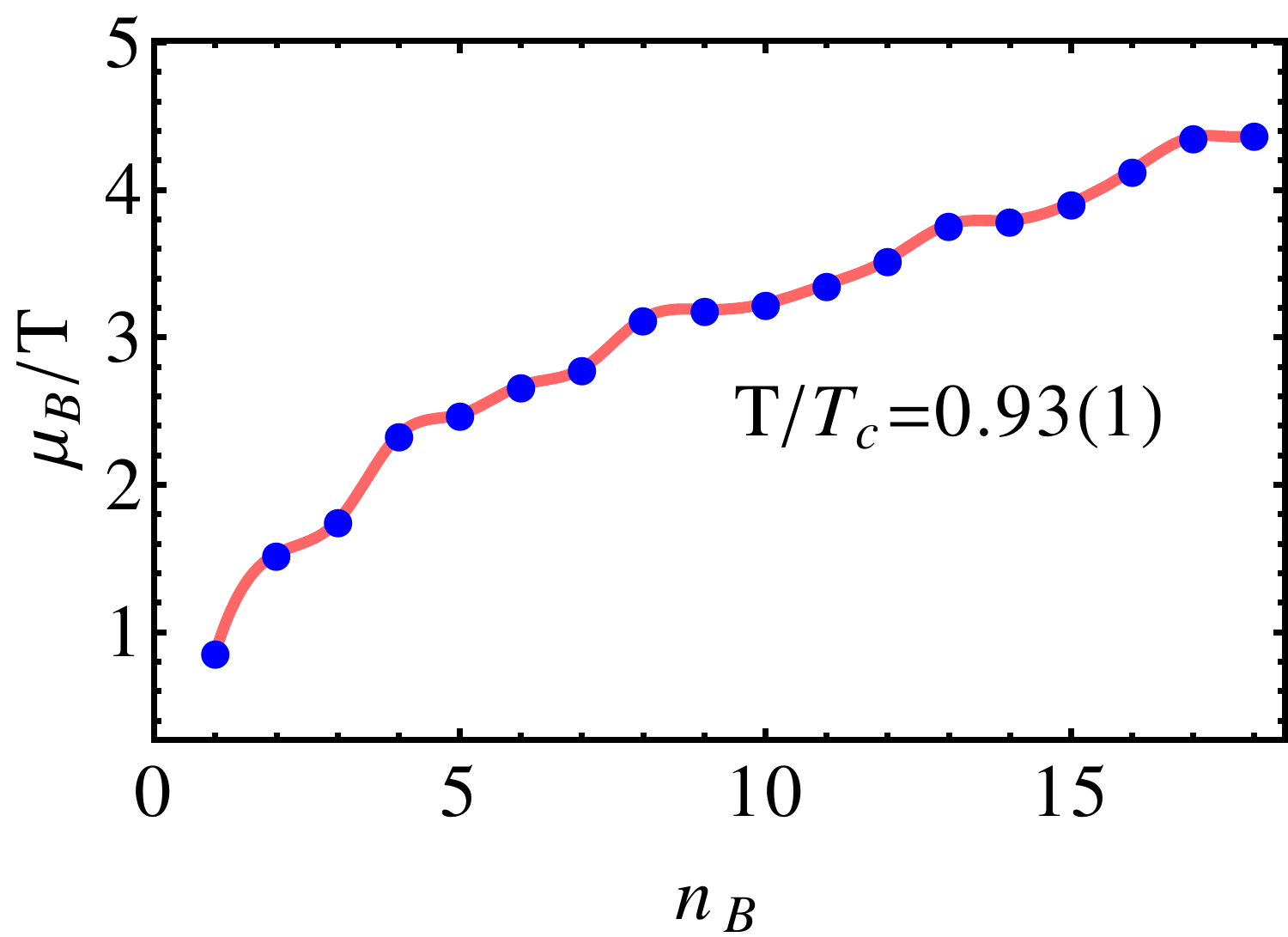}
      \includegraphics[width=1.5in]{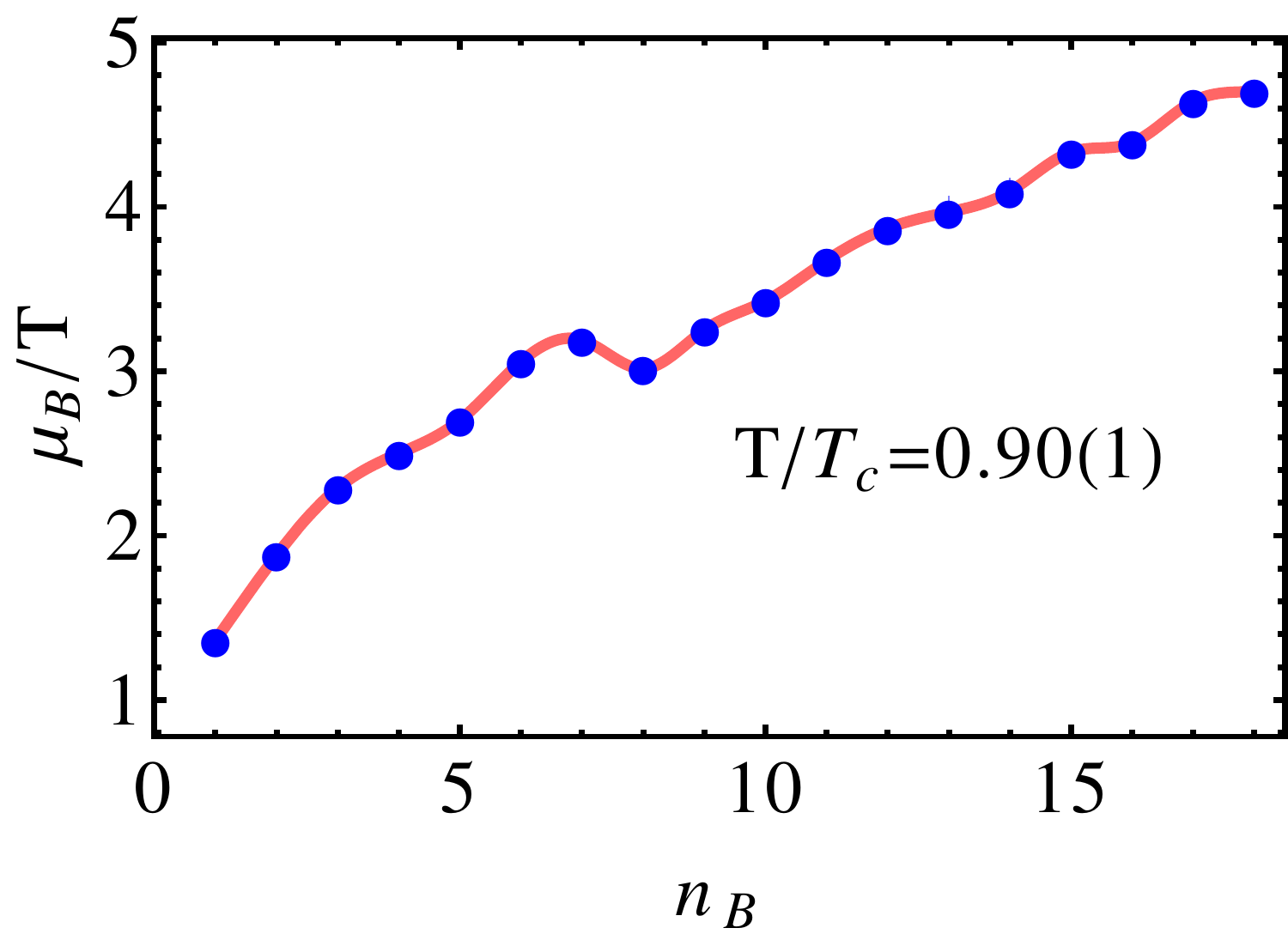}
      \includegraphics[width=1.5in]{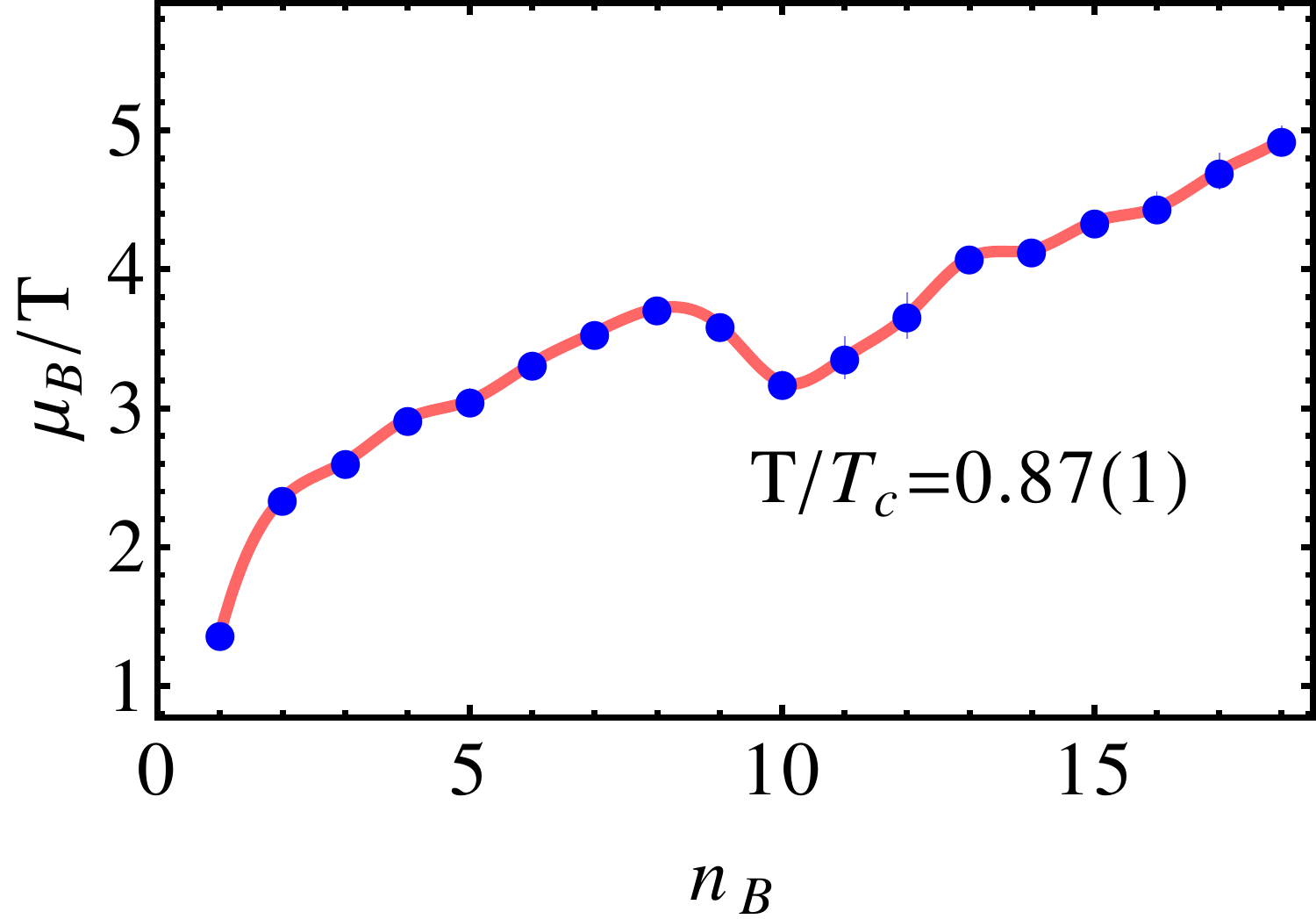}
      \includegraphics[width=1.5in]{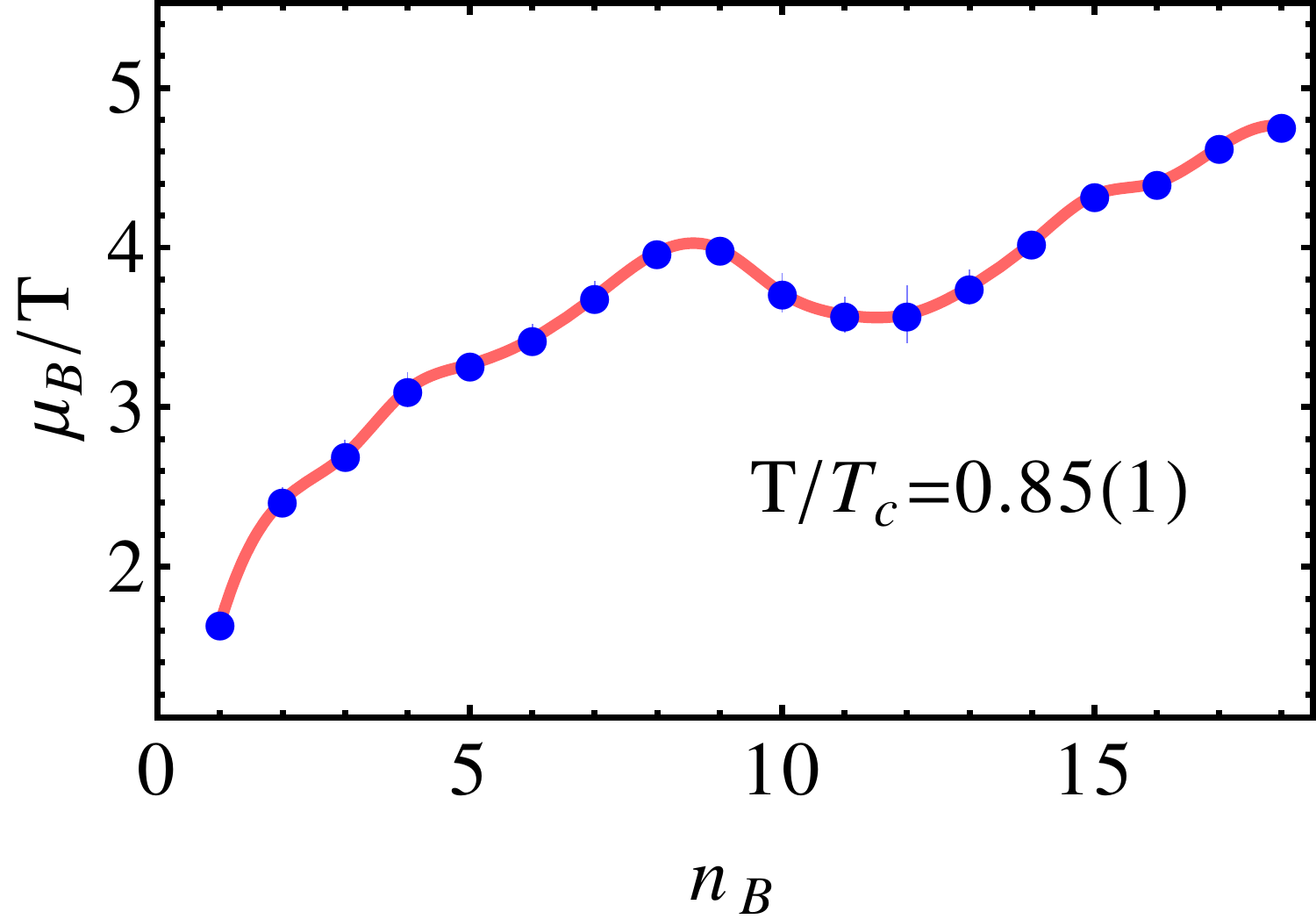}

\caption{Phase scan for temperatures $0.85~T_c$, $0.87~T_c$, $0.90~T_c$ and $0.93~T_c$.}\label{fig:scaning_nf3}
\end{figure}

Our results for the baryon chemical potential are presented in Fig.~\ref{fig:scaning_nf3} for four
different temperatures below $T_c$. Statistical errors are estimated from the jackknife method.
It is clear that the chemical potential exhibits an ``S-shaped'' wiggle for $n_B$ between $6$ and $14$.
To identify the boundaries of the mixed-phase region and the coexistence baryon chemical potential,
we rely on the Maxwell construction: the coexistence chemical potential $\widetilde{\mu_{B}}$ is the
one that produces equal areas between the curve of the chemical potential $\mu_B$ as a function of
$n_B$ and the constant $\widetilde{\mu_B}$ line which intersects with $\mu_B$ at $n_{B_1}$ and $n_{B_2}$.
This procedure was used in studies with staggered fermions~\cite{deForcrand:2006ec,Kratochvila:2005mk}
and Wilson-clover fermions~\cite{Li:2010qf} in this context for the $N_f=4$ case.

\begin{figure}[t]
\includegraphics[width=2.4in]{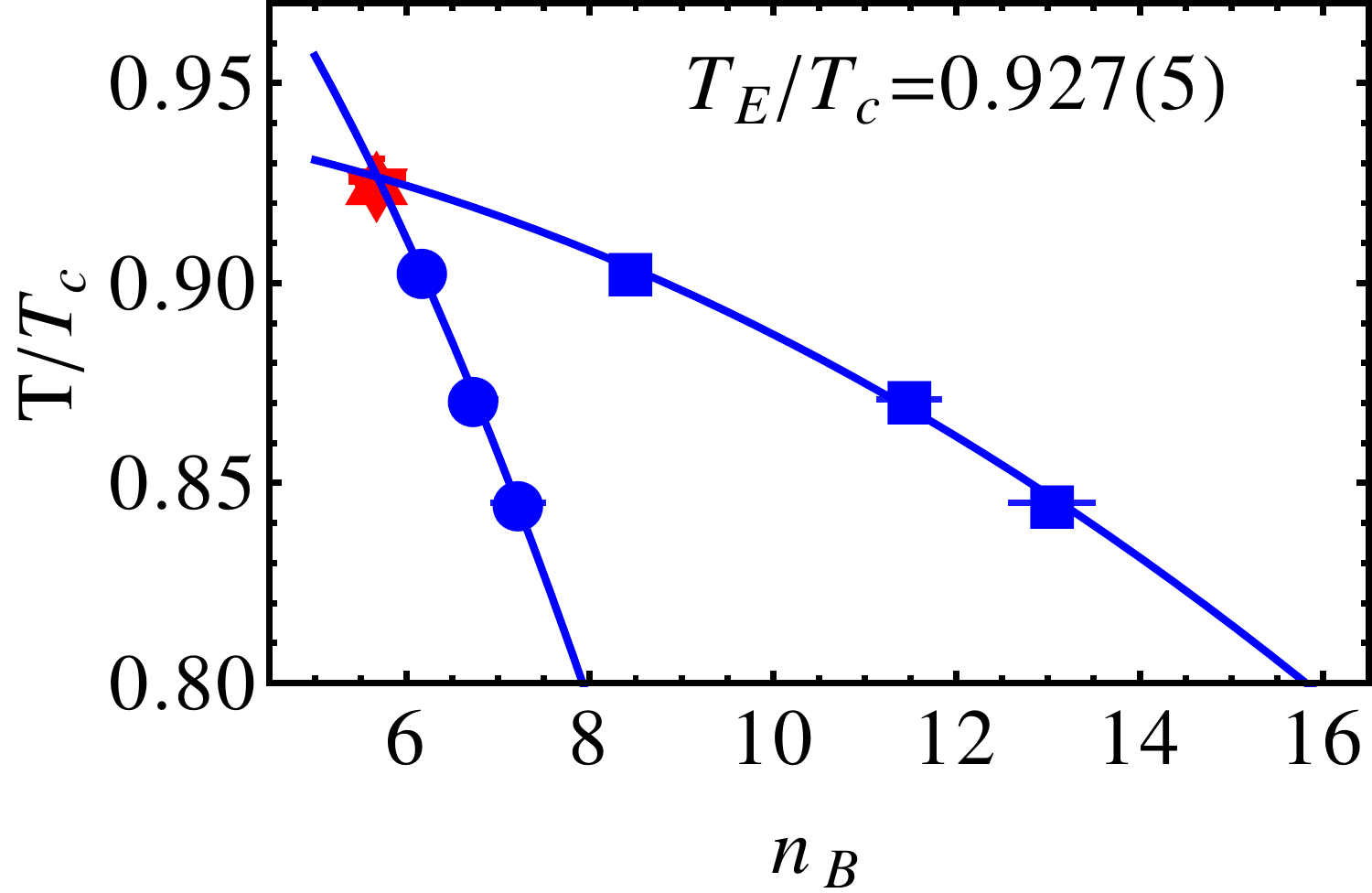}
\caption{Phase boundaries in the canonical ensemble.}\label{fig:t_rho}
\end{figure}

We carried out the Maxwell constructions for the three temperatures at $0.85~T_c$, $0.87~T_c$ and $0.90~T_c$.
We could not do it for the 0.93 $T_c$ case, as the wiggle there, if present, is not statistically significant.
The results are presented in Fig.~\ref{fig:maxwell}. Having determined $n_{B_1}$ and $n_{B_2}$ for three
temperatures, we plot the boundaries of the coexistence region and perform an extrapolation in $n_B$ and
$T$ to locate the intersection of the two boundaries. To determine the crossing point, we
perform a simultaneous fit of the boundary lines using a even
polynomial in baryon density. We use an even
polynomial since $Z_C$ is an even function of $k$.
The phase boundaries and
their extrapolations are plotted in Fig.~\ref{fig:t_rho}. We find the intersection point at
$T_E(n_B^E)/T_c = 0.927(5)$ and $n_B^E=5.7(3)$.

\begin{figure}[t!]
\includegraphics[width=2.5in]{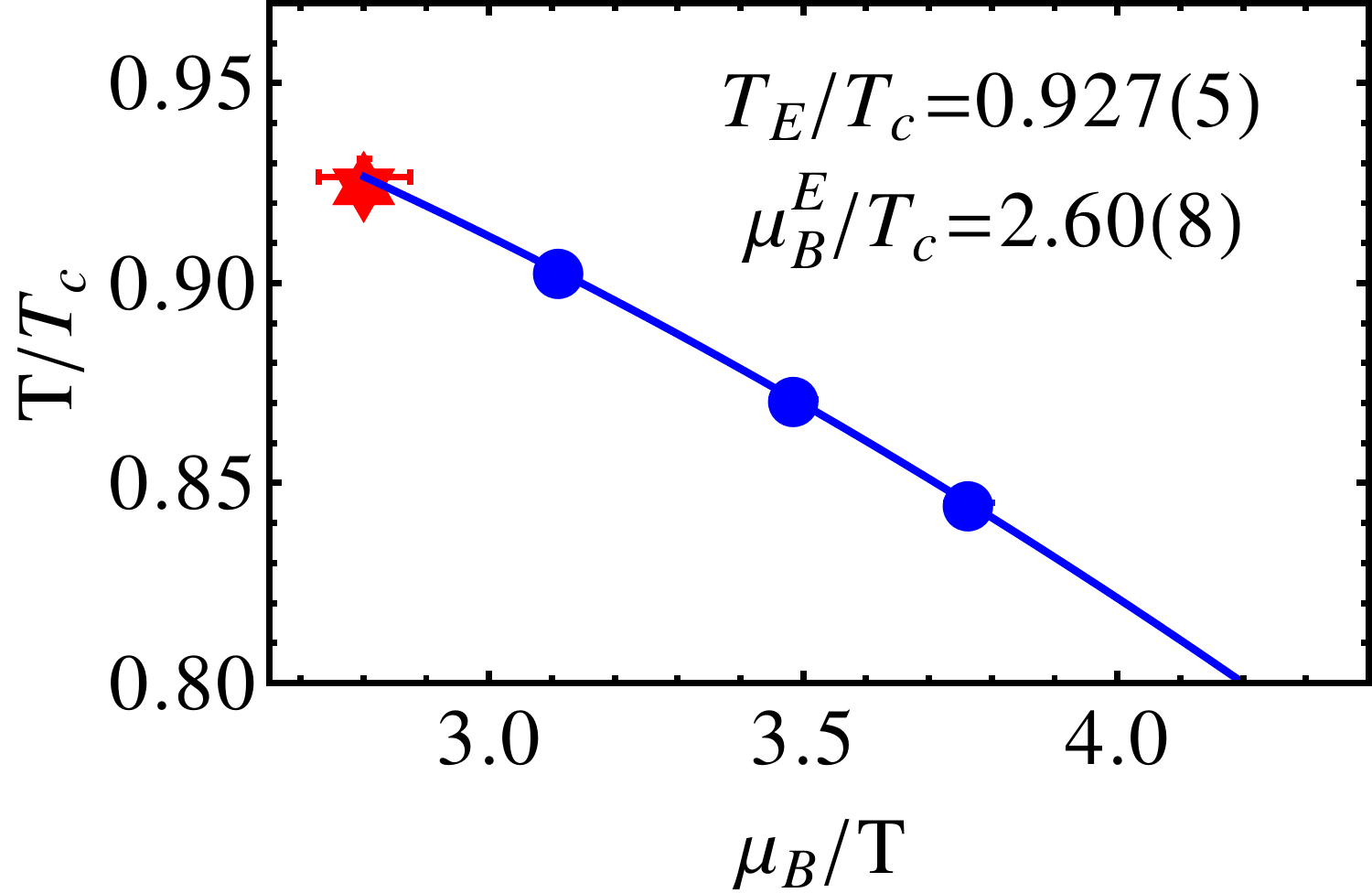}
\caption{Phase transition line in the $T$,
$\mu$ plane.}\label{fig:t_mu}
\end{figure}

Using the coexistence chemical potential, one can map out the phase diagram in the grand canonical
ensemble as shown in Fig.~\ref{fig:t_mu}. Note that, the region of coexistence phase becomes a
curved transition line separating two the phases as we expected. In this way, we locate the critical
point in  the grand canonical ensemble at critical temperature $T_E/T_c = 0.927(5)$ and baryon chemical
potential $\mu_B^E/T_c = 2.60(8)$. Using the lattice spacing $a \approx 0.3\fm$ in our simulation, we
convert its location in physical units to be $T_E \approx 157 \MeV$ and $\mu_B^E \approx 441\MeV$.

In conclusion, we have applied a canonical ensemble algorithm previously tested on the $N_f = 4$ to the more relevant $N_f =3$ case and located the first order phase transition
as signaled by the S-shape structure in the $\mu - n_B$ plane for several temperatures below $T_c$. The Maxwell construction was employed to identify the boundaries of the coexistence phase and we extrapolated them to locate the critical point at $T_E = 0.925(5)~T_c$ and $\mu_B^E = 2.60(8)~T_c$. We should point out that the present work is carried out on a relatively small volume with spatial extent of $\sim 1.8 \fm$ and for three degenerate quark flavors with their masses similar to that of the strange quark. Quark mass for this system acts like the magnetic field for spin systems which weakens the phase transition. Since the $\mu =0$ finite temperature transition is first order for massless quarks~\cite{Pisarski:1983ms} and the present critical point is at a relatively large $\mu_B^E$ for quark masses around the strange, one expects that the critical point for the more realistic $2+1$ flavor case with light $u/d$ quarks to be somewhere in between. This expectation is based on the assumption that there is a critical surface which grows continuously from the critical line at $\mu = 0$ into finite $\mu$. The critical line is the one that separates the first order phase transition region at some finite temperature with small quark masses and the crossover region with intermediate masses (including the physical ones) on the $\mu = 0$ plane of the Columbia plot. This assumption is challenged by recent studies of the critical surface near the critical line (See Fig.1 in references~\cite{deForcrand:2008vr, deForcrand:2008zi}) which suggest that the first order region shrinks with increasing chemical potential and, therefore, there might not be a critical point for physical quark masses. To address this issue, future simulations will study the quark mass dependence of the critical point and the existence of the critical point needs to be checked on lattices with higher cutoffs and larger volumes.

This work is partially supported by DOE grants DE-FG05-84ER40154, DE-FG02-05ER41368 and DE-FG02-95ER-40907. We wish to thank P.~de Forcrand, A.~Kennedy and S.~Chandrasekharan for useful discussions.

\bibliography{myref}
\newpage

\end{document}